\documentclass[
 longbibliography,
superscriptaddress,
 amsmath,amssymb,
 aps,
]{revtex4}

\usepackage{graphicx}
\usepackage{dcolumn}
\usepackage{bm}

\usepackage{natbib}
\begin{document}

\preprint{APS/123-QED}

\title{Label-free quantitative imaging of two-dimensional concentration gradients using Fabry-P\'erot interferometry}

\author{T. Peshkovsky*}
\affiliation{Department of Materials, ETH Z\"{u}rich, Vladimir-Prelog-Weg 1-5/10, 8093, Z\"{u}rich, Switzerland
}
\author{S. A. Schmid*}
\affiliation{Department of Materials, ETH Z\"{u}rich, Vladimir-Prelog-Weg 1-5/10, 8093, Z\"{u}rich, Switzerland
}
\author{D. Taylor}
\affiliation{Department of Mechanical and Process Engineering, ETH Z\"{u}rich, 8092, Z\"{u}rich, Switzerland
}
\author{Robert W. Style}
\affiliation{Department of Materials, ETH Z\"{u}rich, Vladimir-Prelog-Weg 1-5/10, 8093, Z\"{u}rich, Switzerland
}
\author{Lucio Isa}
 \email{lucio.isa@mat.ethz.ch}
\affiliation{Department of Materials, ETH Z\"{u}rich, Vladimir-Prelog-Weg 1-5/10, 8093, Z\"{u}rich, Switzerland
}
\email{lucio.isa@mat.ethz.ch}

\author{Federico Paratore}
\email{federico@unbound-potential.com}
\affiliation{Department of Materials, ETH Z\"{u}rich, Vladimir-Prelog-Weg 1-5/10, 8093, Z\"{u}rich, Switzerland
}

\date{\today}

\begin{abstract}
Concentration gradients at the microscale play a central role in many physical, chemical, and biological systems, yet their quantitative visualization remains challenging due to the limited optical contrast associated with changes in concentration. Here, we present RIO (the Refractive Index Observer), a label-free interferometric tool for quantitative imaging of refractive index, and thus concentration, fields in microfluidic systems. Implemented using a Fabry–Pérot microfluidic chip mounted on a standard optical microscope, RIO achieves a per-pixel refractive index precision on the order of $1 \cdot 10^{-5}$ refractive index units (RIU) using a standard CMOS camera, enabling high sensitivity two-dimensional chemical imaging. We characterize the refractive index resolution and spatiotemporal performance of the instrument and demonstrate its capabilities by measuring concentration gradients of dissolved NaCl in a co-laminar flow. RIO provides an accessible, label-free platform for quantitative studies of microscale concentration fields in systems where molecular labeling is undesirable or impractical, and enables investigations of a broad range of out-of-equilibrium phenomena, from polymerization and enzymatic reactions to cell signaling and electrochemical processes.
\end{abstract}

\maketitle


\section{\label{sec:level1}INTRODUCTION}

The ability to visualize microscale concentration gradients is central to understanding and controlling processes in a wide range of biological, electrochemical, and chemical systems. Despite their importance, direct observation of concentration gradients remains challenging because most liquid systems exhibit only weak optical contrast arising from concentration variations. To date, gradient visualization typically relies on fluorophores or dyes to enhance contrast, making gradients observable under fluorescence or light microscopy. However, labeling can modify the physicochemical properties of the solute and perturb the processes under investigation. In addition, not all compounds can be effectively labeled. Moreover, dyes and fluorophores are prone to photobleaching and, in biological environments, can be cytotoxic (\cite{kim_three-dimensional_2016, shaked_label-free_2023}). These limitations highlight the need for label-free approaches to measure concentration gradients.

An alternative is offered by detection of refractive index contrast as a label-free measure of concentration variations. Qualitative phase imaging techniques that use refractive index differences, such as phase contrast microscopy and differential interference contrast microscopy, are common in various application fields (\cite{shakher_review_1999, agarwal_temperature_2018, mohan_refractive_2019, crespi_three-dimensional_2010, bornhop_free-solution_2007, nowbahar_measuring_2018}). However, refractive index measurements with \textit{quantitative} two-dimensional spatiotemporal resolution are less accessible, typically requiring expensive and dedicated equipment or complex data analysis (\cite{pandey_measurement_2017, zhang_quantitative_2017, rappaz_measurement_2005}). As a result, fluorescence or tracer-based methods are still commonly used, and label-free,  accessible techniques for refractive index mapping remain lacking (\cite{katuri_inferring_2021, hokmabad_chemotactic_2022}).

In 2015, Vogus \textit{et al.} introduced a Fabry–Pérot interferometric system that allows for label-free measurement of concentration gradients in microfluidic channels by monitoring shifts in \textit{Fringes of Equal Chromatic Order} (FECO) arising from refractive index variations (\cite{rvogus_measuring_2015}). Their setup used a monochromatic light source passing through a microfabricated Fabry–P\'erot microfluidic chip, with the transmitted signal analyzed using a spectrometer to resolve local refractive index values, achieving a sensitivity on the order of $10^{-5}$ refractive index units (RIU). In this configuration, they demonstrated spatiotemporal visualization of  \textit{one-dimensional} refractive index variations, i.e. along a single spatial line, which was subsequently applied to investigate diffusion coefficients in aqueous solutions (\cite{rvogus_measuring_2015}) and ionic liquids (\cite{bayles_anomalous_2019}), as well as to study interfacial polymerization kinetics (\cite{nowbahar_measuring_2018}).

Building on the work of Vogus \textit{et al.}, here we present a label-free approach that enables \textit{two-dimensional} visualization of concentration gradients in a Fabry-P\'erot microfluidic chip. We refer to this instrument as the \textit{Refractive Index Observer} (RIO). At the core of RIO is a tunable optical filter that precisely selects the wavelength of light illuminating the microfluidic chip, with an averaged spectral resolution of 0.02 nm, while the resulting FECO are recorded pixel by pixel on a standard camera. This configuration enables the detection of refractive index changes on the order of $1 \cdot 10^{-5}$ RIU per image pixel, with a spatial resolution determined by the microscope magnification and a temporal resolution governed by the wavelength-scanning speed, camera acquisition rate, and associated signal-to-noise ratio.

In the following, we describe RIO’s design and working principle, characterize its resolution, and demonstrate its performance by quantifying the diffusive mixing of an aqueous NaCl solution and deionized water in a co-laminar flow.

\section{\label{sec:level2}WORKING PRINCIPLE}

\begin{figure}[h]
    \centering
    \includegraphics[width=0.9\linewidth]{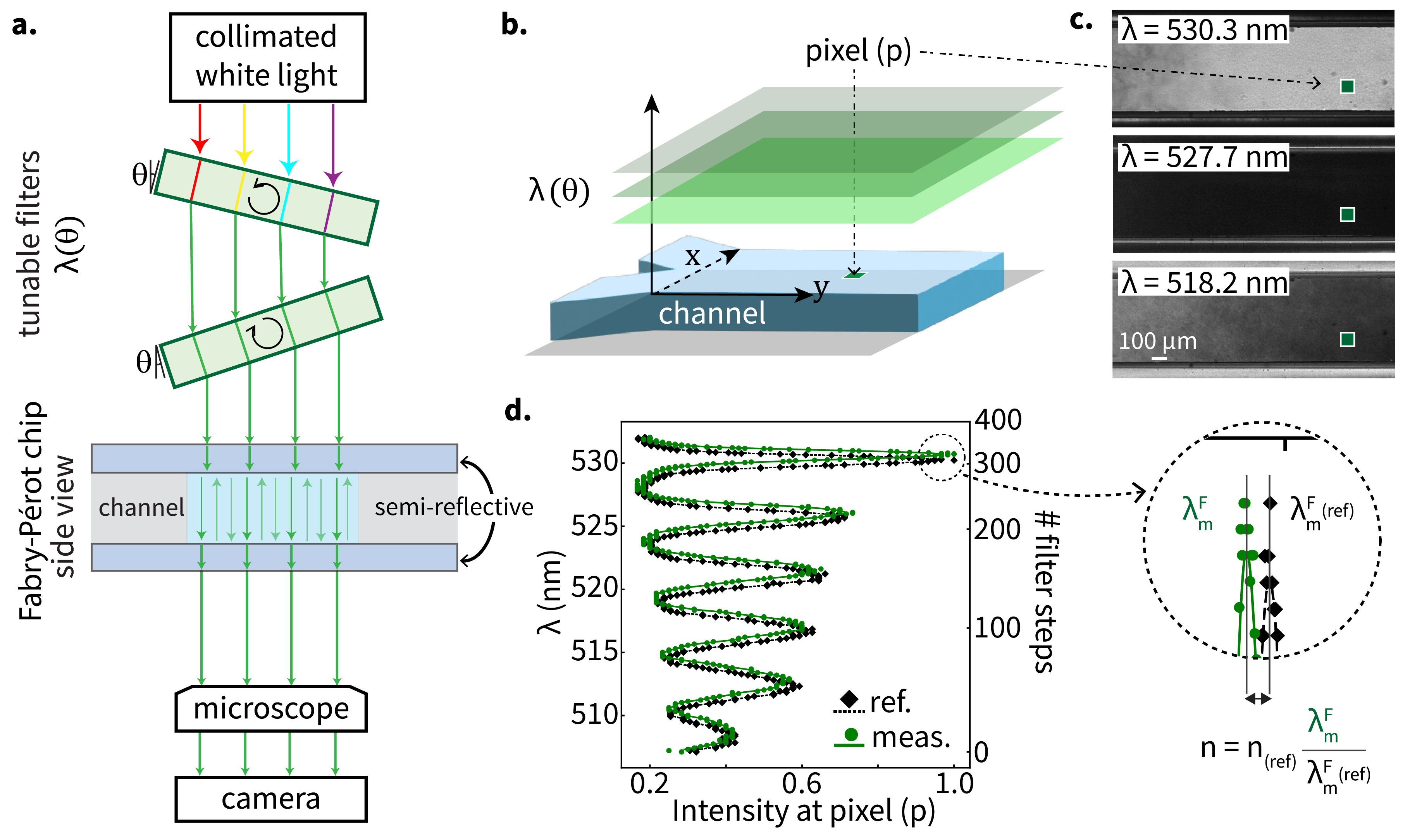}
    \caption{\textbf{(a)} Operating principle of RIO. Monochromatic light, with its wavelength selected by adjusting the angle of two rotating dichroic filters, illuminates a microfluidic Fabry–P\'erot chip. The spatial interference  pattern is imaged through an inverted microscope onto a CMOS camera. One measurement consists of a complete wavelength scan ranging from $\sim 508$ nm to $\sim 532$ nm.
    \textbf{(b)} Illustration of wavelength-resolved image stacking. As the filters rotate, a wavelength $\lambda(\theta)$ is selected and the entire field of view is sequentially illuminated, one wavelength at a time, yielding a total of 400 images per measurement. A representative image pixel $p$ is indicated. 
    \textbf{(c)} Example images of a microfluidic Fabry–P\'erot chip filled with deionized water, acquired at three different wavelengths, with pixel $p$  highlighted.
    \textbf{(d)} Intensity spectra measured at pixel $p$ for an initial reference state (black diamonds) and a subsequent measurement (green circles). The spectra are smoothed with a Gaussian kernel (see SI section 3, figure S1) to identify FECO; the inset highlights the wavelength shift of the first peak. Using the initial refractive index $n_{\text{(ref)}}$, the shifted FECO wavelengths are used to determine the new refractive index $n$ at pixel $p$.
    }
    \label{fig:WorkingPrinciple}
\end{figure}

A schematic of RIO's working principle is shown in figure \ref{fig:WorkingPrinciple}a. Collimated polychromatic light passes through two dichroic band-pass filters mounted on high-precision stepper motors that synchronously adjust their angle of incidence, $\theta$. Because the transmitted wavelength depends on $\theta$, sweeping $\theta$ enables scanning across different wavelengths.  The first filter provides a coarse spectral selection with a full-width at half-maximum (FWHM) of $50$ nm at a center wavelength of $\lambda_0 = 535$ nm, while the second filter has an ultra narrow bandpass with a FWHM of only $0.17$ nm at a center wavelength of $\lambda_0 = 532.08$ nm to selectively bin wavelengths (see Materials and Methods, section \ref{Materials}). 

The selected monochromatic light passes through the Fabry–Pérot microfluidic chip, which is custom-fabricated from borosilicate glass and coated on both internal sidewalls with a semi-reflective silver layer (see SI sections 1 \& 2). The chip acts as an optical modulator with a relationship between the transmitted intensity ($I_t$) and the incident intensity ($I_0$) described by the Airy function 

\begin{equation}\label{eq:Airy}
I_t=I_0\left[\frac{1}{1+\left(\frac{2 R}{1-R^2}\right)^2 \sin ^2\left(\frac{2 \pi n d}{\lambda}\right)}\right],
\end{equation}

\noindent where $R$ is the reflectivity of the walls, $\lambda$ the incident wavelength, $d$ the distance between the semi-reflective walls, and $n$ is the local refractive index between them (\cite{hariharan_basics_2010}). At resonant wavelengths, when the incident light and reflected light constructively interfere, $I_t$ is at its maxima.  

By scanning $\lambda$, we record the transmitted intensity $I_t(\lambda)$ at each pixel across the entire field of view using a CMOS camera installed on a standard microscope (an image of the complete experimental set-up can be found in figure S2). This yields wavelength-resolved images of the sample, as illustrated in figures \ref{fig:WorkingPrinciple}b and \ref{fig:WorkingPrinciple}c. A single sweep in $\lambda$ produces a stack of images that together constitute one measurement. For each pixel, the resulting $I_t(\lambda)$ curve exhibits a characteristic signal (figure \ref{fig:WorkingPrinciple}d) in which the FECO peak wavelengths are 

\begin{equation}\label{eq:fringe}
    \lambda_m^F = \frac{2nd}{m},
\end{equation}

\noindent where $m$ is the chromatic order of the fringe. 

RIO operates in two modes that differ in the quantity being measured: an \textit{absolute method} that measures the refractive index itself, and a \textit{relative method} that tracks refractive index changes. In both cases, the measurement error is fundamentally limited by the precision of wavelength selection, such that  $\frac{\delta n}{n} \propto \frac{\delta \lambda}{\lambda}$ (\cite{rvogus_measuring_2015}). Using the \textit{absolute method}, the refractive index is derived by

\begin{equation}\label{eq:absMeas}
n=\left(\frac{1}{2 d}\right) \frac{\lambda_m^{\mathrm{F}} \lambda_{m+1}^{\mathrm{F}}}{\lambda_m^{\mathrm{F}}-\lambda_{m+1}^{\mathrm{F}}}, 
\end{equation}

\noindent which relies on knowing the thickness of the channel, $d$, for any pixel in the field of view. In principle, this can be achieved by measuring $d$ using a medium of known refractive index with the same equation \ref{eq:absMeas}; however, this calibration step propagates additional uncertainty into the refractive index measurement, as the error for absolute measurements scales as  $\frac{\delta n}{n} \propto\frac{d}{\lambda} \frac{\delta \lambda}{\lambda}$. In addition, the \textit{absolute method} relies on measuring the wavelengths of  adjacent FECO peaks, $\lambda_m^{\mathrm{F}}$, whose uncertainty depends on the precision in selecting $\theta$. Because the relation between $\delta \theta$ and transmitted wavelength $\delta \lambda$ is non-linear (see SI section 5, figure S3), the wavelength resolution is most accurate, i.e., $\delta \lambda\approx 0.16 pm$, at normal incidence ($\theta \sim 0^\circ$), and worsens at higher angles (e.g., $\delta \lambda\approx0.1 nm$ at $\theta \sim \pm 36^\circ$). Consequently, the uncertainty of the \textit{absolute method} is limited by the precision of determining different FECO peaks far from the center wavelength  $\lambda_0 = 532.08$. For our system, this limits the resolution of the \textit{absolute method} to approximately $\sim 10^{-3}$ RIU.  

Alternatively, the \textit{relative method} focuses on tracking the shift of one selected \textit{m-th} FECO wavelength between different measurements. The relation between the refractive index of an initial medium $n_{\text{(ref)}}$ and the one of a subsequent measurement $n$ is  

\begin{equation}\label{eq:relMeas}
n = n_{\text{(ref)}} \frac{\lambda_m^F}{\lambda^F_{m\text{(ref)}}},
\end{equation} 

\noindent By tracking the FECO closest to the central wavelength, $\lambda_0 = 532.08$, as shown in figure \ref{fig:WorkingPrinciple}d, we maximize the precision in wavelength selection, resulting in a resolution on the order of $1 \cdot 10^{-5}$ RIU. Hence, throughout this work we employ the \textit{relative method}. 

\section{\label{sec:level3}TOOL CHARACTERIZATION}
\begin{figure}[h!] 
    \centering
    \includegraphics[width=0.9\linewidth]{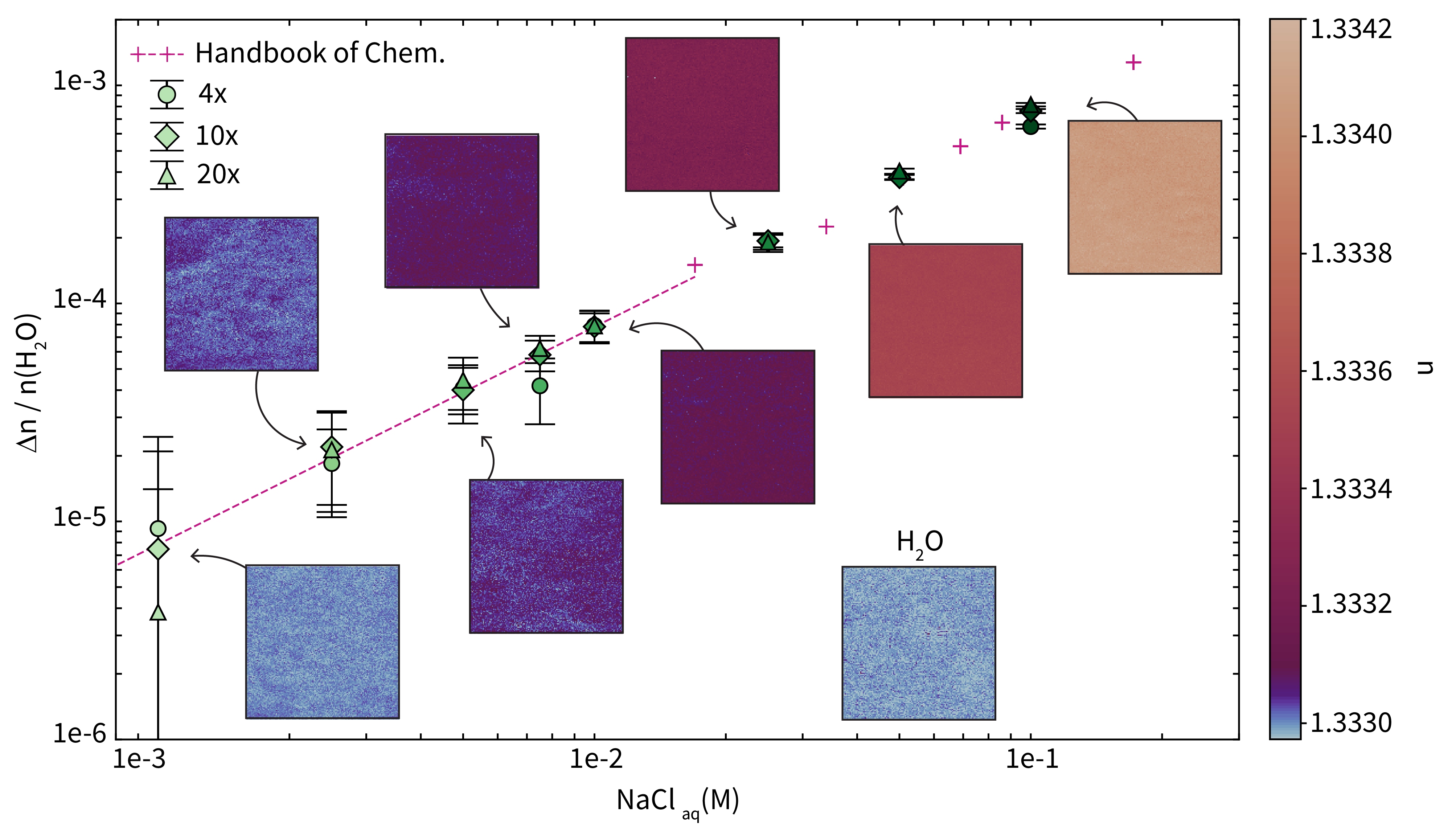} 
    \caption{
    Normalized refractive index shift of aqueous NaCl solutions with concentrations ranging from 1 mM to 100 mM measured with RIO at 4x, 10x, and 20x magnifications. Error bars represent the standard deviation of the refractive index measured over a 10 x 10 pixel area. Reference values from the \textit{Handbook of Chemistry and Physics 100th Edition} (\cite{rumble_crc_2019}) are shown as (+) signs, and a linear projection of these values is plotted as a dashed line for concentrations below 10 mM. Representative 20x magnification RIO measurements for each NaCl concentration are shown alongside the corresponding data points, with colors corresponding to the refractive index according to the color-bar on the right. A RIO image of deionized water is also shown for comparison. 
    }
    \label{fig:PrecisionPlot}
\end{figure}

\textbf{Refractive index resolution.} To characterize the resolution of RIO, we measured the refractive index of aqueous NaCl solutions at concentrations ranging from 1 mM to 100 mM. Figure \ref{fig:PrecisionPlot} shows the normalized refractive index differences between each NaCl solution and deionized water ($\Delta n$ / $n_{(H_{2}O)}$), as a function of NaCl concentration. The calculation of $n$ values from the RIO measurements is shown in figure S4. The data correspond to the average of eight consecutive measurements with identical parameters (see figure S5 for additional details on how to improve the signal-to-noise ratio). The refractive index resolution of RIO, estimated by averaging the spatial standard deviation of the calculated refractive index across all concentrations, is on the order of  $1 \cdot 10^{-5}$ RIU. This corresponds to one $\sigma$ of the measured RI value distribution in a homogeneous medium. 

To validate the accuracy of RIO, the measured refractive index values were compared to reference values reported in the CRC Handbook of Chemistry and Physics 100th edition (\cite{rumble_crc_2019}), which are also plotted in figure \ref{fig:PrecisionPlot}. The close agreement between the two datasets demonstrates both the accuracy of the refractive index measurements and the absence of biases over the investigated concentration range. 

As a reference, a typical benchtop refractometer used for bulk, point-wise refractive index measurements (e.g. Abbemat Refractometer 500) has a resolution of approximately $1 \cdot 10^{-5}$ RIU. This comparison shows that the refractive index resolution achieved by  RIO for individual pixels is on par with that of a conventional bulk refractometer, which reports a single value for an entire sample volume.   

Given the sensitivity of the instrument, one needs to eliminate potential thermal effects induced by illumination during measurements. To test this aspect, we performed long-term stability measurements at a controlled environmental temperature over a period of 18 hours, confirming the intrinsic stability of the system and the absence of measurable temperature increases due to illumination (see SI section 8, figure S6). Temperature variations may nevertheless arise from fluctuations in the external environment during prolonged measurements or from exothermic or endothermic processes within the medium under study. Such temperature changes can affect both the intrinsic refractive index of the measured solution, as well as the Fabry–P\'erot height through thermal expansion of the SU8 spacer forming the microfluidic channel. To quantify this dependence, we performed temperature cycling experiments on a chip filled with air, varying the temperature from 23°C to 29°C over 14 hours (see SI section 9, figures S7 \& S8). Our results show that for SU8 spacers, thermal expansion and the resulting change in the channel height $d$, constitutes the dominant contribution to the temperature-induced refractive index variations in air, offering a precise way to calibrate the measurements in fluids exhibiting a stronger T-dependence of refractive index. These findings highlight the importance of accounting for and characterizing possible thermal effects on the refractive index measurements.

\textbf{Spatiotemporal resolution.} RIO is mounted on a standard optical microscope, and its spatial resolution is defined by the optics of the instrument. In particular, we demonstrated the use of 4x, 10x and 20x objectives (0.13, 0.3, and 0.45 numerical apertures respectively) to image NaCl solutions at different concentrations. Figure \ref{fig:PrecisionPlot} shows that measurements acquired at different magnifications overlap within uncertainty, indicating that the refractive index resolution of RIO is independent of the magnification used. At 4x magnification the field of view is 4.2 mm x 4.2 mm with a lateral pixel size of 7.3 $\mu$m, whereas at 20x magnification is 0.8 mm x 0.8 mm with a lateral pixel size of 1.43 $\mu$m. This flexibility enables RIO to visualize refractive index gradients over a range of spatial scales. Representative images of refractive index profiles at these different magnifications are shown in figure S9. 

In addition to defining the spatial sampling, the choice of magnification also affects the temporal resolution of RIO, as higher magnification reduces the photon flux per pixel and therefore requires longer exposure times to maintain a sufficient signal-to-noise ratio. In our current set up, each refractive index measurement is reconstructed from 400 wavelength-resolved images using pixel binning, resulting in total acquisition times ranging from 8 s at 4× magnification to 24 s at 20× magnification.

\section{\label{sec:level4}DIFFUSION MEASUREMENTS}

\begin{figure}[h!] 
    \centering
    \includegraphics[width=0.9\linewidth]{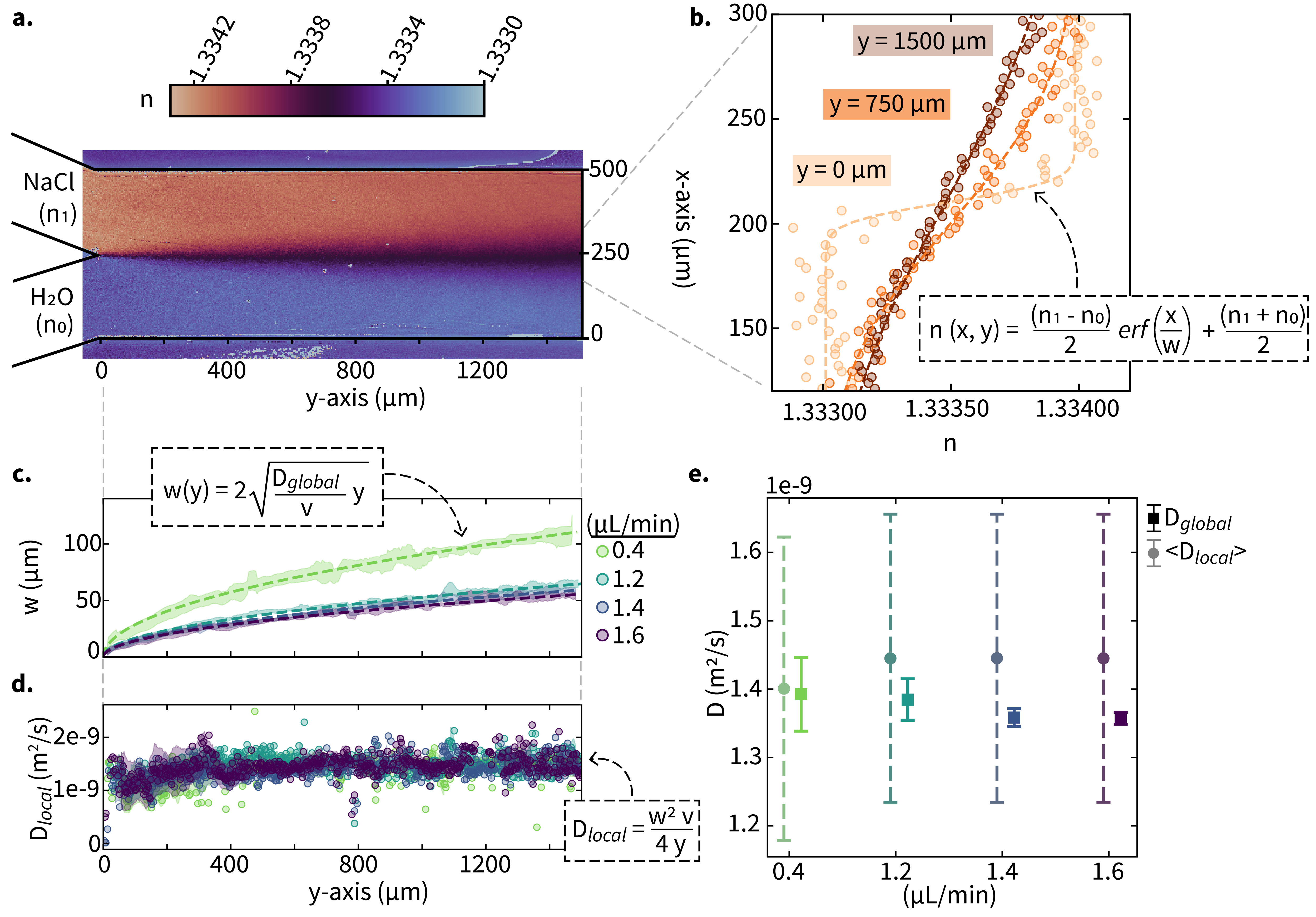}
    \caption{\textbf{(a)} Refractive index map of 100mM NaCl solution (top, n$_{1}$) and deionized water (bottom, n$_{0}$) streams co-flowing in a y-junction channel at $0.4$ $\mu$L/min from left to right. \textbf{(b)} Refractive index as function of the $x$-coordinate at positions $y = 0$ $\mu m$, $y = 750$ $\mu m$, and $y = 1500$ $\mu m$ extracted from the RIO image in \textbf{a} and fitted using equation \ref{eq:diffequn}. \textbf{(c)} Mixing layer width $w(y)$ obtained using the fitting equation \ref{eq:diffequn} and plotted against the $y$-position. The shaded region is the standard deviation of 4 consecutive measurements. Four different co-laminar flow rates were tested: $0.4$ $\mu$L/min, $1.2$ $\mu$L/min, $1.4$ $\mu$L/min, and $1.6$ $\mu$L/min. \textbf{(d)} The diffusivity $D_{local}$ is evaluated from every discrete "$w$" value along the $y$-axis using equation \ref{eq:IWeqnD}. The circle marker is the mean of 4 consecutive measurements, and the shaded region represents their standard deviation. \textbf{(e)} Comparison of $D_{global}$ and the mean $<D_{local}>$ at different flow rates. $D_{global}$ is evaluated by fitting the curves of $w(y)$ shown in \textbf{c} using equation \ref{eq:IWeqn}, where $<D_{local}>$ is the average of the values plotted in \textbf{d} representing the diffusivity extracted using the 1D information at individual $y$-locations along the channel.
    }
    \label{fig:Diffusion}
\end{figure}

As a demonstration of the power of detecting two-dimensional refractive index maps, we report the use of RIO to visualize and quantitatively characterize the diffusion of a model system composed of NaCl in deionized water, showing that the two-dimensional data afford a much higher precision in measuring diffusion coefficients compared to one-dimensional line scans. As shown in figure \ref{fig:Diffusion}a, we establish a co-laminar flow in a Y-shaped microfluidic channel, where a stream of deionized water containing 100 mM of NaCl flows alongside a stream of deionized water. Figure \ref{fig:Diffusion}b shows examples of refractive index profiles, $n(x, y)$, measured across the concentration gradient generated by the diffusion of NaCl along the $x$-direction for three downstream positions. The diffusivity can be estimated directly from the measured refractive index profiles by fitting them to a two-dimensional analytical advection–diffusion model. This approach relies on several assumptions and considerations regarding the experimental system. 

First, we performed the experiments at low P\'eclet numbers, i.e. Pe$_H = (v \cdot H) / D \ll 70$, where $v$ is the average fluid velocity, $H$ is height of the channel, and $D$ is the diffusion coefficient; in this regime, shear-induced dispersion via the Taylor–Aris mechanism is negligible, and mixing between the two streams is governed predominantly by molecular diffusion (\cite{aris_dispersion_1956, vedel_transient_2012}). 

Second, we assume the flow to be steady, fully developed, and predominantly unidirectional along the channel axis, such that transverse advection can be neglected. We also assume that the two inlet streams are homogeneous across the channel height and that any residual concentration gradients across the vertical $z$ direction are rapidly suppressed by diffusion, with a characteristic time scale of $\sim 200$ $msec$ (for a channel height of $20$ $\mu m$ and diffusivity of $\sim 10^{-9}$ $m^2/s$). This time scale is much shorter than the other relevant time scales in the systems, allowing us to treat molecular transport as two-dimensional in the $x-y$ plane. 

Furthermore, we require the mixing layer width $w(y)$ to remain small compared to the channel width so that wall effects can be neglected and we can assume an infinite-domain in the $x$-direction. In our experiments, the maximum observed mixing width is approximately 100 $\mu m$ at the downstream edge of the field of view, well below the total channel width of 500 $\mu m$. Finally, we assume that the refractive index varies linearly with NaCl concentration within the 0-100mM range, as shown in figure \ref{fig:PrecisionPlot}, and that the diffusivity is constant over the investigated concentration range.

Under these assumptions, we can fit the profiles $n(x,y)$ using the analytical solution of the two-dimensional advection-diffusion equation for an initially sharp interface and infinite domain in the $x$-direction,

\begin{equation}\label{eq:diffequn}
n(x,y) = \frac{(n_1 - n_0)}{2} \operatorname{erf}\left(\frac{x}{w(y)}\right) + \frac{(n_1 + n_0)}{2},
\end{equation}

\noindent to extract values of $w(y)$. Here, $n_0$ and $n_1$ are the refractive indices of the pure water and the 100 mM NaCl streams, respectively. 

In the following, we describe two approaches that use the measured $w(y)$ to evaluate the molecular diffusivity. The first approach measures $w$ at discrete downstream positions across the field of view and fits the resulting data to extract the diffusivity using

\begin{equation}\label{eq:IWeqn}
w(y) = 2 \sqrt{\frac{D_{global}}{v} (y + y_0)} ,
\end{equation}

\noindent where $y_0$ is the first contact point of the fluids, $v$ is the mean velocity of the fluid and $D_{global}$ is the diffusivity, and the subscript \textit{global} indicates that it is measured from the change of $w(y)$ along the whole visible part of the channel. The results of this \textit{global} approach are shown in figure \ref{fig:Diffusion}c, where the dashed line represents the fit, and the data are filtered to retain only values with a standard deviation within the inter-quartile range. We ran experiments at 0.4, 1.2, 1.4, and 1.6 $\mu$L/min flow rates, and as expected, increasing the flow rate results in a narrower mixing region. We also compared the measured values with simulations performed using Comsol (Multiphysics 6.3 Win ML), showing a good agreement (see SI section 11, figure S10).

The second approach is to evaluate the diffusivity at each downstream position using the equation

\begin{equation}\label{eq:IWeqnD}
D_{local} = \frac{w^{2} \cdot v}{4 \cdot (y + y_0)}.
\end{equation}

\noindent Here the subscript $local$ reflects that this diffusivity is evaluated independently at each $y$-position. The results of this method are illustrated in figure \ref{fig:Diffusion}d, where only data points with a standard deviation within the inter-quartile range are shown in order to filter out $D_{local}$ values affected by local channel non-uniformity. The extracted diffusivities exhibit negligible dependence on flow rate over the range investigated. This one-dimensional analysis is analogous to the information typically obtained from line-scanning diffusion measurements (\cite{rvogus_measuring_2015}). 

A comparison between diffusivities obtained using the \textit{local} one-dimensional approach and the \textit{global} two-dimensional approach is shown in figure \ref{fig:Diffusion}e. For the mean $D_{local}$, the uncertainty reflects the spatial variations in the measured refractive index (on the order of $10^{-10}$ m$^2$/s), whereas the uncertainty in $D_{global}$ reflects inter-measurement variation (on the order of $10^{-11}$ m$^2$/s). It is worth noting that while the uncertainty associated with $D_{local}$ remains consistent across the different tested flow rates, the uncertainty in $D_{global}$ increases at lower flow rates. We attribute this behavior to the precision of the syringe pump controlling the flow. Although all tested flow rates exceed the pump’s minimum pulsation-free operating limit, lower flow rates are still more susceptible to flow inaccuracies.

Consequently, when diffusivity is extracted using the two-dimensional fitting approach, i.e. using $D_{global}$, the dominant source of uncertainty shifts from the spatial refractive index resolution of the measurement to the stability of the experimental flow conditions. A comparison of the two methods demonstrates that leveraging the full two-dimensional information available in RIO measurements significantly improves diffusivity resolution.

\section{\label{sec:level5}CONCLUSIONS AND OUTLOOK}

In this work, we presented RIO, a label-free interferometric tool for quantitative imaging of refractive index in microfluidic systems. By using a Fabry–Pérot microfluidic chip, we achieved spatially resolved refractive index maps with a per-pixel precision on the order of $1 \cdot 10^{-5}$ RIU, comparable to that of a typical bulk benchtop refractometer. Importantly, RIO can be mounted directly onto a standard optical microscope and does not require dedicated or expensive equipment. We validated the performance of RIO by measuring NaCl concentration gradients in co-laminar flow and by extracting molecular diffusivity with a resolution comparable to state-of-the-art techniques.

While RIO produces extremely accurate refractive index maps, its temporal resolution can be further improved to push it towards resolving fast dynamic processes. At present, it takes a few seconds to acquire a full RIO image and the temporal resolution limited by the signal-to-noise ratio of the wavelength-resolved measurements and by the need to scan multiple wavelengths per acquisition. Several modifications could substantially improve the time resolution. For instance, increasing the transmission of the Fabry–P\'erot microfluidic chip—currently on the order of 1\%—would allow for shorter exposure times while maintaining sufficient signal-to-noise ratio. This could be achieved by reducing the thickness of the reflective layers or by using materials with lower optical absorption. In addition, restricting the wavelength scan to a single FECO peak or reducing the number of wavelength steps would directly decrease acquisition times, enabling access to faster transient phenomena.

Additionally, the refractive index resolution can be improved even further. In fact, the sensitivity of RIO is ultimately determined by the precision of wavelength selection. While the current implementation already achieves a resolution on the order of $1 \cdot 10^{-5}$ RIU, further improvements are feasible through more precise angular control of the tunable optical filters, for example by employing motors with finer step sizes or closed-loop angular encoders. Such improvements would directly translate into enhanced refractive index resolution.

Beyond the specific diffusion experiments presented here, RIO provides a flexible platform for studying microscale transport processes. In particular, it allows direct visualization of out-of-equilibrium phenomena such as diffusion–reaction coupling, interfacial polymerization, and electrochemical processes in application fields ranging from cell-to-cell communications and soft-materials to battery processes and active materials.

\section{\label{Materials}MATERIALS AND METHODS}
\textit{Light Source:} A solid-state white light (Lumencor light engine, SOLA III 6-LCR-NK) is used as the illumination source. It has an output power of 3.4 W and is connected to RIO via a 3 mm diameter liquid light guide (Thorlabs, LLG3-6H). The white light then goes through a collimator and a pinhole to prevent the formation of ghost fringes in the Fabry P\'erot chip. Optical alignment was performed to ensure that the light is sufficiently collimated (see SI section 12).

\textit{Optical Filters:} Two optical dichroic filters in series function as the tunable wavelength selection mechanism; one filter acts as a wavelength selector while the other filter compensates for the beam offset generated by the other filter. The first filter (Alluxa, 535-50 OD6) has a FWHM of $50nm$ at a peak wavelength at normal incidence of $\lambda_0 = 535nm$. The second filter (Alluxa, 532.08-0.17 OD5)  has an ultra narrow bandpass with a FWHM of $0.17nm$ at a peak wavelength at normal incidence of $\lambda_0 = 532.08nm$. The dichroic filters do not change their bandpass linearly with respect to the angle of incidence (see SI section 5, figure S3).Thus, when the incident light is normal to the filters a single motor step of 0.09$^{\circ}$ results in a wavelength increment on the order of $0.16 pm$, while at the largest AOI of 36$^{\circ}$ the wavelength increment is in the order of $0.1 nm$. 

\textit{Stepper Motors:} Each optical filter is rotated by a stepper motor (Nanotec, PD2-C4118L1804-E-01) which has 4096 steps per revolution and a stopping accuracy of 0.09$^{\circ}$. The stepper motors are in sync with the camera acquisition and controlled via an Arduino micro-controller; a detailed connection scheme is shown in SI section 13, figure S11.

\textit{Microscope:} RIO is mounted on an inverted microscope (Nikon ECLIPSE Ti2). The 4x, 10x, and 20x magnification objectives with NA of 0.13, 0.3, and 0.45 respectively were used. 

\textit{Camera:} A digital CMOS camera (Hamamatsu ORCA-Fusion C14440-20UP) with 2304 x 2304 pixels records RIO measurements. 

\textit{Fabry-P\'erot microfluidic chip:} The Fabry-P\'erot chip doubles as both an optical cavity and a microfluidic chip. Two borosilicate glass wafers are coated with a thin 50 nm layer of Ag that serves as the semi-reflective layer. The Ag layer is sandwiched between an adhesive 2 nm Ti layer, and a protective 10 nm SiO$_{2}$ layer. Channels are formed with $20$ $\mu$m of SU8 patterned using standard lithography on one glass wafer which is then bonded to the another coated glass wafer. A schematic of the microfluidic chip is shown in figure S2. For the details regarding Fabry-P\'erot fabrication see SI section 2.

\textit{NaCl Experiments:} To characterize the spatiotemporal resolution and RI precision of RIO we used a series of concentration gradients formed at interface of a colaminar flow of pure Milli-Q water and aqueous NaCl (Sigma-Aldrich) solutions of different concentrations. The flow rates are controlled by using two syringe pumps (CETONI GmbH, NEM-B101-03 A) connected with 0.51 mm ID tubing (Masterflex™ Tygon™ Microbore Tubing) and interfaced with the microfluidic chip using a 4-ways Dolomite microfluidic connector.

\subsection*{Abbreviations}
\begin{tabular}{ l l  }
 {\bf RI} & refractive index \\ 
 {\bf FECO} & fringes of equal chromatic order  \\  
 {\bf RIU} & refractive index units  \\
 {\bf RIO}  & refractive index observer \\
 {\bf FWHM} & full-width half maximum \\
 {\bf AOI} & angle of incidence

\end{tabular}

\begin{acknowledgments}
We gratefully acknowledge Miguel Angel Chamorro Burgos from the University of Seville for fruitful discussions. The authors also gratefully acknowledge the clean-rooms FIRST at ETH Zurich and the Binnig and Rohrer Nanotechnology Center at IBM Zurich for their support and assistance in this work. The authors acknowledge the use of ChatGPT by OpenAI for editorial assistance, including language polishing and typographical corrections.\newline
\end{acknowledgments}

\emph{Patent application:}
A patent on RIO is pending. 
\newline

\emph{Funding Statement:}
F.P. acknowledges the financial support from the Swiss National Science Foundation (Ambizione Grant No. 209056). The authors acknowledge ETH Zurich for financial support.
\newline

\emph{Declaration of Interests:}
The authors declare no conflict of interest.
\newline

\emph{Author Contributions:}
F.P., L.I., and D.T. designed the research. D.T. and F.P conceptualized RIO. F.P., D.T., and S.S. built the first prototype. S.S. characterized RIO and performed preliminary experiments. T.P. performed the research. R.S. and T.P. analyzed the diffusion experiments. L.I. provided funding. L.I. and F.P. supervised the project. T.P. wrote the original manuscript with input from all authors. F.P., L.I., and T.P. edited the manuscript.  
\newline

\emph{Data Availability Statement:}
Raw data are available from the corresponding authors. 
\newline

\emph{Ethical Standards:}
The research meets all ethical guidelines, including
adherence to the legal requirements of Switzerland.


\end{document}